\documentclass[3p,times,twocolumn]{elsarticle}

\usepackage{ecrc}

\volume{00}

\firstpage{1}

\journalname{Nuclear and Particle Physics Proceedings}

\runauth{P.~Ru and B.-W.~Zhang}

\jid{nppp}

\jnltitlelogo{Nuclear and Particle Physics Proceedings}

\usepackage{amssymb}

\usepackage[figuresright]{rotating}

\begin{document}

\begin{frontmatter}



\dochead{}

\title{Probing cold nuclear matter effects with weak gauge boson production in ultra-relativistic heavy-ion collisions}

\author[label1,label2]{Peng Ru}
\author[label2]{Ben-Wei Zhang}
\address[label1]{School of Physics $\&$ Optoelectronic Technology, Dalian University of Technology, Dalian 116024, China}
\address[label2]{Key Laboratory of Quark \& Lepton Physics (MOE) and Institute of Particle Physics,
 Central China Normal University, Wuhan 430079, China}



\begin{abstract}
Within the framework of pQCD, we systematically study the $W^{\pm}/Z^0$ boson production as a probe of cold nuclear matter effects and nuclear parton distributions in p+Pb and Pb+Pb collisions at the LHC and future colliders.
A detailed analysis at partonic level is performed.
Moreover, with a semi-microscopic KP model of NPDFs, in which several nuclear effects (e.g. Fermi motion and nuclear binding, the off-shell correction, the nuclear coherent correction, and the nuclear meson correction) are included, we study the vector boson rapidity distribution in p+Pb collisions at the LHC, and a very good agreement with the latest data is found, including the $W$-boson charge asymmetry.

\end{abstract}

\begin{keyword}
weak gauge boson \sep cold nuclear matter effects \sep nuclear PDFs \sep W charge asymmetry

\end{keyword}

\end{frontmatter}


\section{Introduction}
\label{Introduction}
The cold nuclear matter~(CNM) effects or the nuclear parton distribution functions~(NPDFs) are quite important for the
hard-scattering processes in high-energy nuclear collisions.
They also provide the baseline for the study of the quark-gluon-plasma~(QGP) property with hard probes
in relativistic nucleus-nucleus collisions~\cite{Albacete:2013ei, Albacete:2016veq}.
Largely impeded by the non-perturbative mechanism therein, it is hard to fully compute the nuclear parton distributions from first principle.
Conventionally, the NPDFs are extracted from global fits to the experimental~(e.g. DIS, DY) data.
With great efforts, several groups working on that ~\cite{Eskola:2009uj, deFlorian:2011fp, Schienbein:2009kk} have achieved significant progresses.
However the error bars of the obtained NPDFs and the differences among the CNM effects extracted by different groups are still considerable.

\begin{figure}[h]
\centering
\includegraphics[scale=0.46]{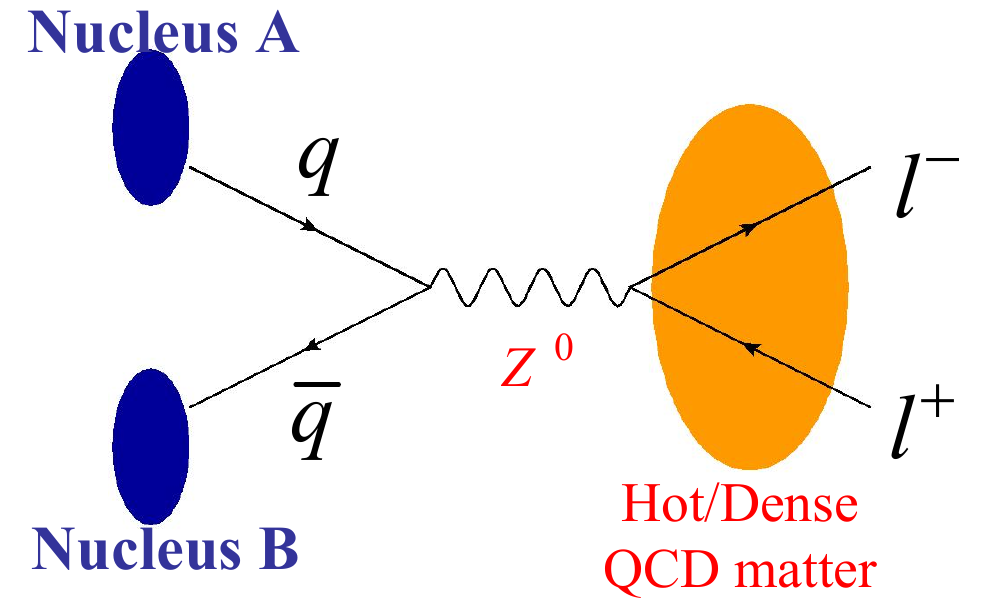}
\caption{A schematic diagram for $Z^0$-boson production through the Drell-Yan~(DY) mechanism in nucleus-nucleus collisions.}
\label{DY}
\end{figure}

With the running of the LHC, the production of the weak gauge boson~($W^\pm/Z^0$)
in heavy-ion collisions provides a new excellent probe of the CNM effects.
With the high-invariant mass~($\sim80-90$~GeV), weak bosons will be produced in the very
early stage~($\sim1/m_{W/Z}\sim10^{-3}$~fm/c) of the collisions,
and can decay later~($\sim0.08-0.09$~fm) to a colorless lepton pair in the
final state~(Drell-Yan), which carries a clean signal of the initial state~(e.g PDFs)~\cite{ConesadelValle:2009vp}.
Even in the nucleus-nucleus collision~(Fig.~\ref{DY}), the created hot and dense QCD medium will hardly pollute this signal.
The weak boson production in nuclear collisions will open a unique opportunity to study the NPDFs at high $Q^2$~($\sim100^2$~GeV$^2$).

In QCD perturbation theory, the cross section of the DY process in hadronic collision can be written
as the convolution of the parton distribution and the partonic hard-scattering cross section,
according to the factorization theorem.
As the factorization is expected to hold in nuclear collisions, the cross section can be calculated by
utilizing the nuclear PDFs.
In this work, we study the vector boson production in nuclear collisions at NLO and NNLO
with the program DYNNLO~\cite{Catani:2007vq,Catani:2009sm} by incorporating nuclear parton
distributions.

\section{Vector boson in nuclear collisions at the LHC}
\label{LHC}
\begin{figure}[t]
\includegraphics[scale=0.6]{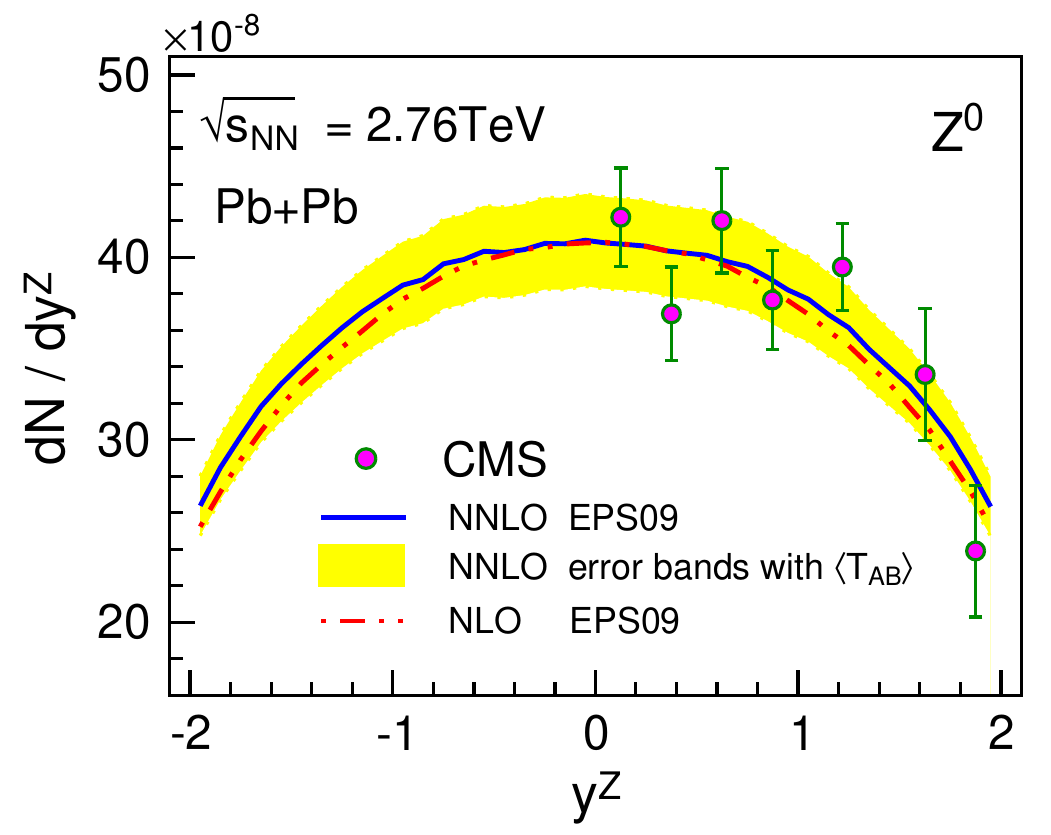}
\caption{The $Z^0$-boson rapidity distribution in Pb+Pb collisions at $\sqrt{s_{NN}}=2.76$~TeV, calculated at NNLO and NLO with EPS09 nuclear modifications.
The experimental data are taken from Ref.~\cite{Chatrchyan:2014csa}. More details can be seen in Ref.~\cite{Ru:2014yma}.}
\label{Z-PbPb-y}
\end{figure}

In this section we focus on the vector boson production in nuclear collisions at the LHC.
We show in Fig.~\ref{Z-PbPb-y} the $Z^0$ rapidity distribution in Pb+Pb collisions at $\sqrt{s_{NN}}=2.76$~TeV.
Both the results at NLO and NNLO with EPS09 NPDFs agree well with the CMS data,
and the NNLO correction is rather small in the studied rapidity region.
In Ref.~\cite{Ru:2014yma}, the NLO result is also found to well describe the $Z^0$ transverse momentum distribution
in large-$p_T$~($\gtrsim 10$~GeV) region in Pb+Pb collisions.

\begin{figure}[t]
\includegraphics[scale=0.65]{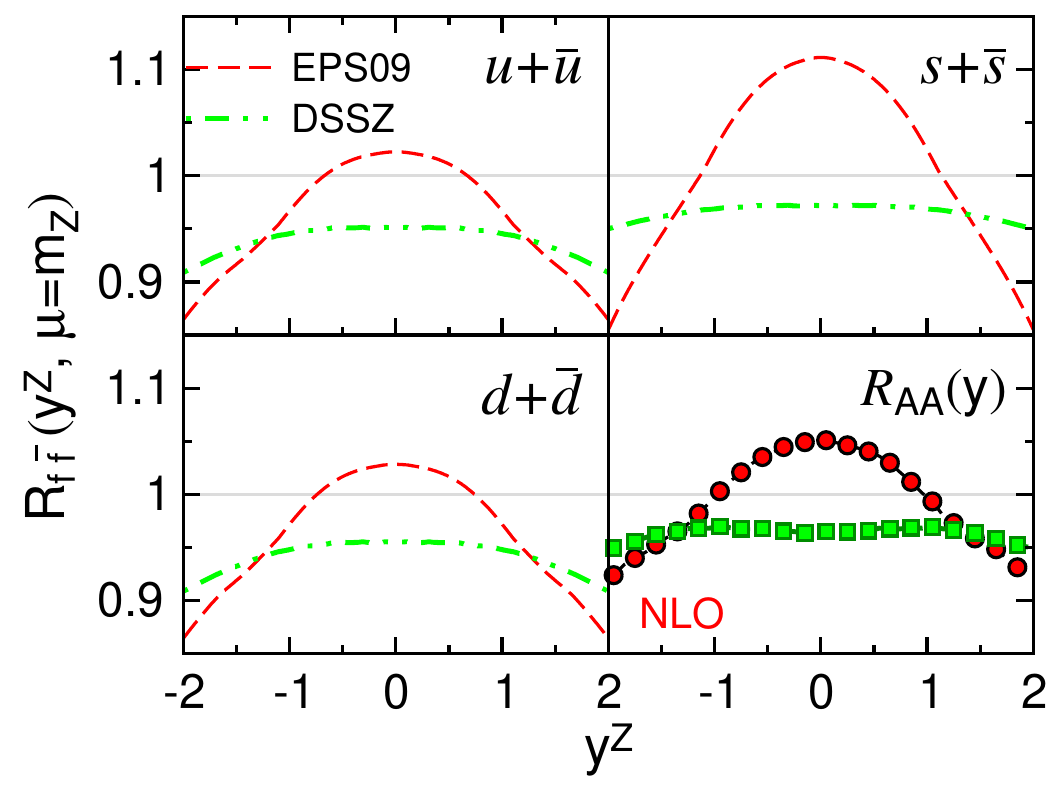}
\caption{Partonic nuclear modification factor $R_{f\bar{f}}(y^Z)$ for different flavors,
and the $R_{AA}(y^Z)$ for $Z^0$ rapidity distribution at NLO~(the last panel), in Pb+Pb collisions at $\sqrt{s_{NN}}=2.76$~TeV.
More details can be seen in Ref.~\cite{Ru:2014yma}.}
\label{Z-Raa}
\end{figure}

In the last panel of Fig.~\ref{Z-Raa}, we show the nuclear modification factor $R_{AA}(y)$ corresponding to Fig.~\ref{Z-PbPb-y}.
It is observed that the predictions with EPS09 and DSSZ NPDF sets show obvious difference, especially in the mid-rapidity region.
This can be understood by performing a partonic level analysis.
The momentum fraction carried by the initial parton can be estimated at LO as $x_{1,2}=m_Ze^{\pm y^Z}/\sqrt{s_{NN}}$.
At the LHC, the kinematic region related to the mid-rapidity corresponds to
the transition between the valence-dominated and the sea-quark-dominated regions~($x\sim0.033$).
Since gluons do not give the LO contribution, the boson rapidity distribution
is much sensitive to the valence and sea quark distributions.
In Fig.~\ref{Z-Raa} we show the partonic nuclear modification factor $R_{f\bar{f}}(y^Z)$
simply defined as $\left[\frac{1}{2}R_f(x_1)R_{\bar{f}}(x_2)+\frac{1}{2}R_f(x_2)R_{\bar{f}}(x_1)\right]$
by considering the LO process $q\bar{q}\rightarrow Z^0$, where the
$R_f(x)\equiv f^{Pb}(x)/f^p(x)$ is the nucelar modification ratio given by NPDFs and the LO relation between $y^Z$ and $x_{1,2}$ is used.
The $R_{AA}$ show similarity with the $R_{f\bar{f}}$ and is an excellent probe of the CNM effects on valence and sea quark distributions.
However we note the precision of the measurement until now is not sufficient to exclude neither EPS09 nor DSSZ.
\begin{figure}[h]
\includegraphics[scale=0.65]{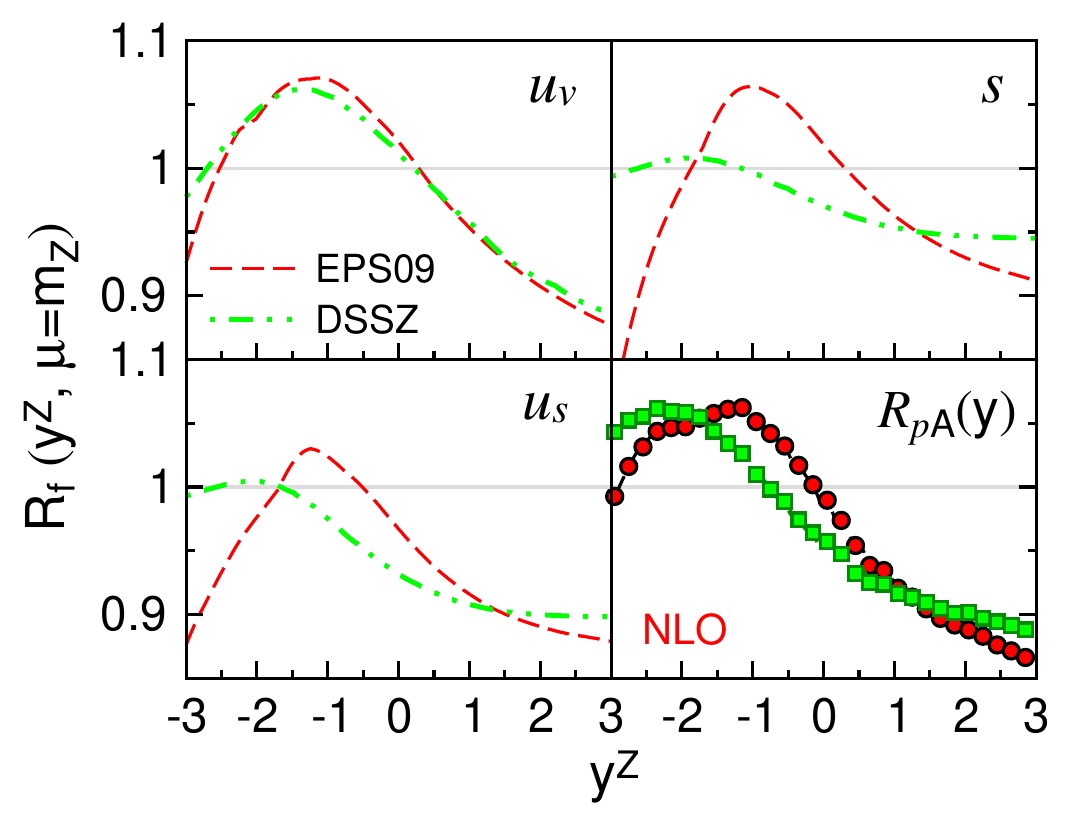}
\caption{Partonic nuclear modification factor $R_f(y^Z)$ for different flavors,
and the $R_{pA}(y^Z)$ for $Z^0$ rapidity distribution at NLO~(the last panel), in p+Pb collisions at $\sqrt{s_{NN}}=5.02$~TeV.
More details can be seen in Ref.~\cite{Ru:2014yma}.}
\label{Z-Rpa}
\end{figure}

\begin{figure}[t]
\includegraphics[scale=0.62]{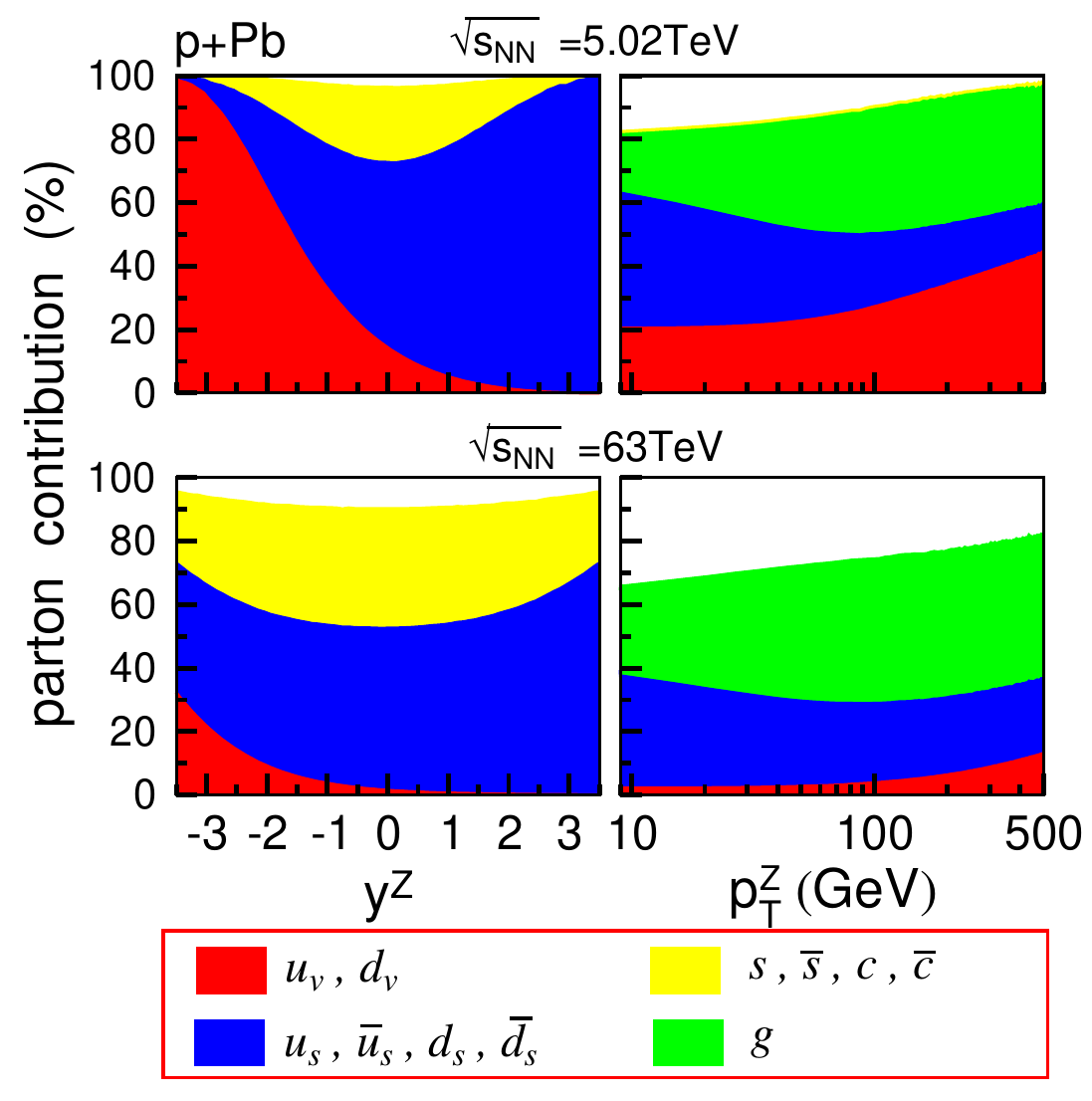}
\caption{Nuclear parton contributions to the $Z^0$ boson rapidity~(left) and transverse momentum~(rght) distributions, in p+Pb collisions at $\sqrt{s_{NN}}=5.02$~TeV~(LHC, top panels) and $\sqrt{s_{NN}}=63$~TeV~(future, bottom panels), calculated at LO. More details can be seen in Ref.~\cite{Ru:2015pfa}.}
\label{Parton-Contribution}
\end{figure}

In p+Pb collisions, the CNM effects usually result in an asymmetric nuclear modification
on $Z$-boson rapidity distribution, as is shown in the last panel of Fig.~\ref{Z-Rpa}.
The partonic level analysis is easier than that in Pb+Pb collisions, since the nuclear partons come from only one direction.
In Fig.~\ref{Z-Rpa} we show the partonic factor $R_f(y^Z)\equiv R_f(x_{Pb})$ given by NPDFs.
We found $R_{pA}(y^Z)$ provides an image of the nuclear modifications on the valence and sea quark distributions.
The suppressions in the forward region is mainly the result of the shadowing effect on the sea quark distributions,
while the enhancements in the backward is largely due to the anti-shadowing effect on the valence quark distributions.
See also the parton contributions in the first panel of Fig.~\ref{Parton-Contribution}.

The nuclear modification on $Z^0$-boson $p_T$ spectrum in heavy-ion collisions at the LHC is also studied~\cite{Ru:2014yma}.
Here we just emphasize that, compared to the rapidity distribution, the $p_T$ spectrum will provide more information
on the nuclear gluon distribution since the gluon can give the lowest-order contributions~(e.g. $qg\rightarrow Z^0q$).
See also the right-top panel in Fig.~\ref{Parton-Contribution}.

Besides the similar nuclear modifications as observed in $Z^0$ production,
those on $W^\pm$-boson production show some unique characteristics, due to the
flavor dependence~($u\bar{d}\rightarrow W^+$, $d\bar{u}\rightarrow W^-$).
For example the strong isospin effect~\cite{Ru:2014yma} and the sensitivity to
the flavor dependent nuclear modifications~(e.g. $r_u\neq r_d$).

\section{Vector boson production in future HIC}
\label{FCC}
In future HIC with much higher energies~\cite{Ru:2015pfa}, the vector boson production will probe CNM effects in new kinematic regions.
The momentum fraction carried by the nuclear partons will decrease a lot.
The shadowing effects on sea quarks and gluons would be more important, while the valence quark contribution and the isospin effect should be small.
See also the parton contributions for $Z^0$ rapidity and $p_T$ distributions in p+Pb collisions at both LHC and future colliders in Fig.~\ref{Parton-Contribution}.
By combining the LHC and the future measurements, we may give new powerful constraints on the NPDFs~\cite{Ru:2015pfa}.
\begin{figure}[t]
\includegraphics[scale=0.62]{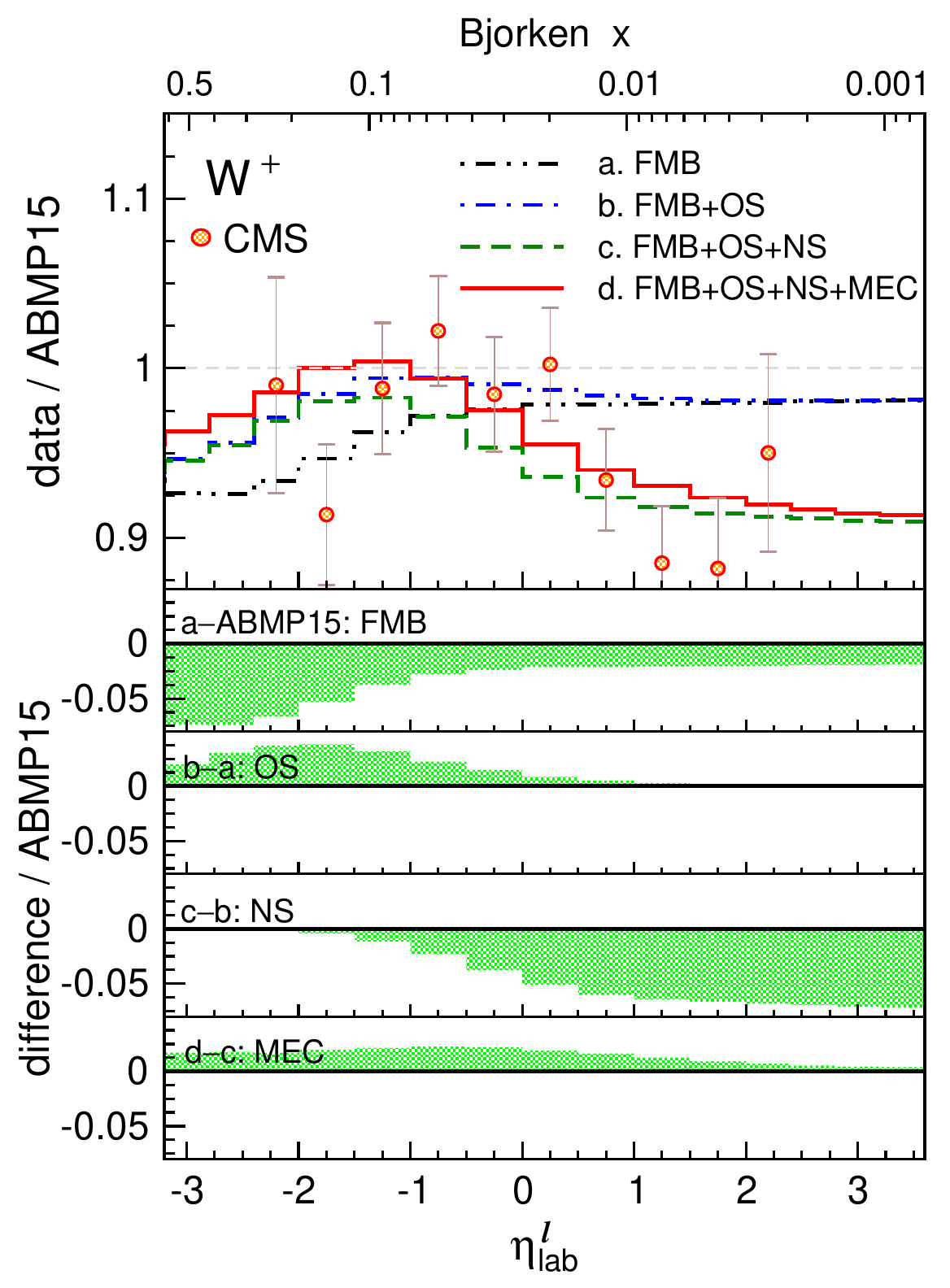}
\caption{Top panel: nuclear corrections on charged-lepton pseudo-rapidity distribution with different
combinations of nuclear effects in KP model with respect to the prediction without nuclear modification~(ABMP15),
calculated at NLO for $W^+$ production in p+Pb collisions at $\sqrt{s_{NN}}=5.02$~TeV.
The CMS data are taken from Ref.~\cite{Khachatryan:2015hha}.
The momentum fraction carried by the nuclear parton estimated at LO are shown on the top axis.
Bottom panels: contributions to the nuclear correction from each individual nuclear effects.
More details can be seen in Ref.~\cite{Ru:2016wfx}.}
\label{W-pPb-eta}
\end{figure}

\section{W/Z production in p+Pb with KP NPDFs}
\label{KP}
\begin{figure}[t]
\includegraphics[scale=0.62]{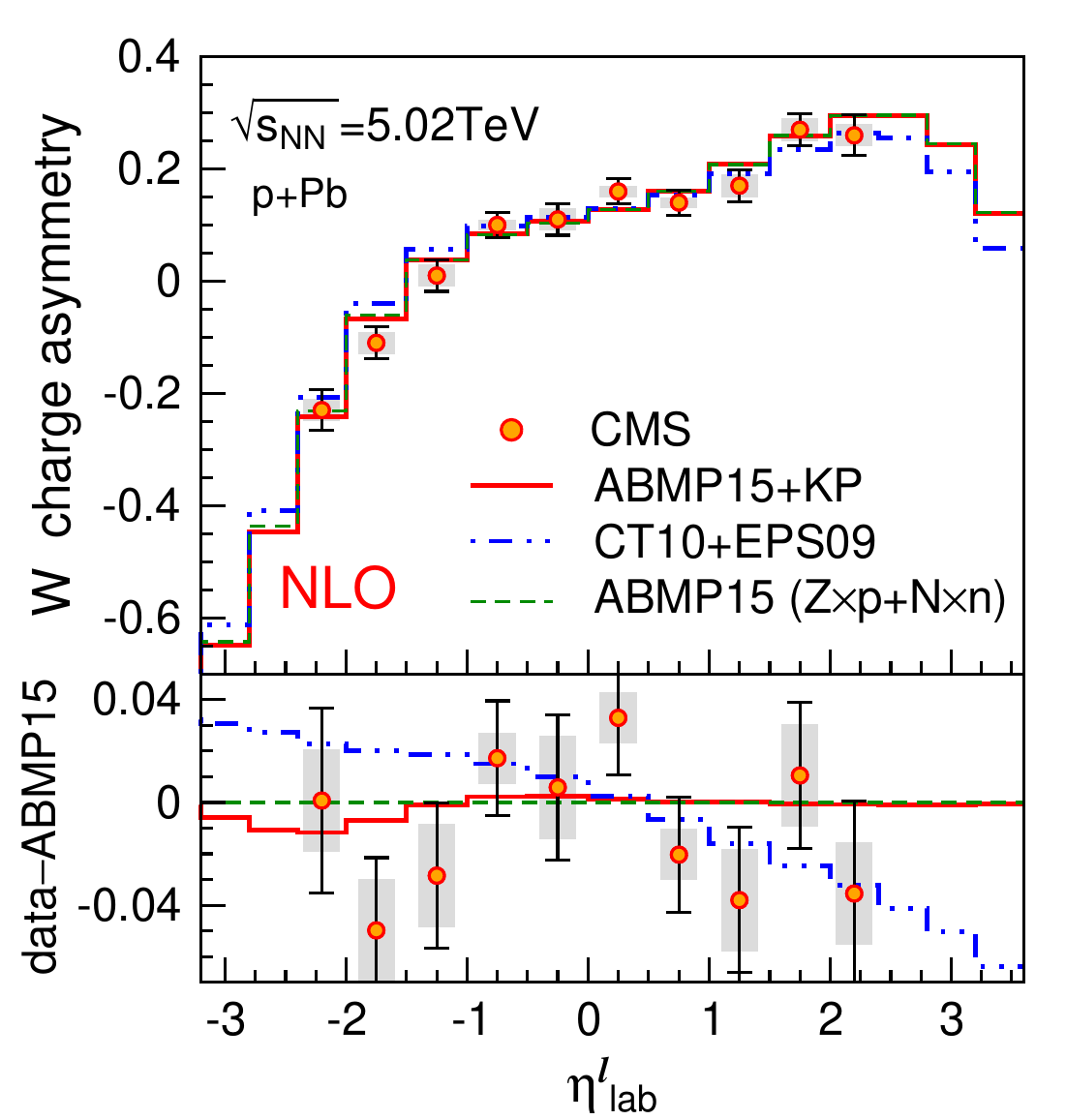}
\caption{Top panel: the $W^\pm$-boson charge asymmetry as a function of the charged-lepton pseudo-rapidity in p+Pb collisions at $\sqrt{s_{NN}}=5.02$~TeV.
Bottom panel: differences between each result and that without nuclear modification~(ABMP15).
The CMS data are taken from Ref.~\cite{Khachatryan:2015hha}. More details can be seen in Ref.~\cite{Ru:2016wfx}.}
\label{W-cas}
\end{figure}

Different from the conventional NPDFs obtained by fitting data, a semi-microscopic-model
based NPDFs have been recently developed by S. A. Kulagin and R. Petti~(KP)~\cite{Kulagin:2004ie,Kulagin:2014vsa}.
The KP NPDFs have been validated with the data from a wide range of processes~(e.g. DIS, DY), and the
offered insights on the underlying nuclear physics mechanisms make it more valuable.
In KP model, different nuclear effects on the parton distribution are taken into account:
the Fermi motion and nuclear binding~(FMB), the off-shell effect~(OS), the coherent nuclear interaction
related to the nuclear shadowing~(NS) and the nuclear meson exchange current~(MEC) correction.

In Ref.~\cite{Ru:2016wfx} we study the rapidity dependence of vector boson production in p+Pb collisions at the LHC with KP NPDFs.
The NLO results with KP NPDFs is found to show an excellent agreement with the latest data from both CMS and ATLAS.
Much meaningfully, we study how each individual CNM effect play a role on
W/Z production in p+Pb collisions at the LHC~(see Fig.~\ref{W-pPb-eta}), through which the experimental data are better understood.
It is found that the full nuclear correction in the rapidity region
measured by CMS and ATLAS is the result of an interplay of different mechanisms.

In Fig.~\ref{W-cas}, we show the interesting result for W-boson charge asymmetry.
It is found the prediction with KP NPDFs show a good agreement with the CMS data.
At the same time we note the difference between the predictions with KP and EPS09 is
largely due to their corresponding proton PDFs~(ABMP15~\cite{Alekhin:2015cza} and CT10~\cite{Gao:2013xoa}).
It is also noteworthy that for the observable sensitive to the CNM effects rather than the proton PDFs,
e.g. the $Z^0$ forward-backward asymmetry, the results with KP NPDFs show a good performance with the CMS data\cite{Ru:2016wfx}.

\section{Conclusion}
We systematically study the $W^{\pm}/Z^0$ boson production as a probe of the CNM effects in
nuclear collisions at the LHC and future colliders, within the framework of pQCD.
Differences among various state-of-art NPDFs sets are found in several observable,
and a partonic-level analysis is performed.
With the KP NPDFs we show how the underlying nuclear physics mechanisms
play a role on the vector boson rapidity distributions in p+Pb collisions at the LHC.
The predictions with KP NPDFs show an excellent performance with latest LHC p+Pb data.

We thank E.~Wang, W.-N.~Zhang and L.~Cheng, S.~A.~Kulagin and R.~Petti for the valuable collaborations. This research is supported by the MOST in China under Project Nos. 2014CB845404, 2014DFG02050, and by NSFC of China with Project Nos. 11322546, 11435004, and 11521064.




\nocite{*}
\bibliographystyle{elsarticle-num}
\bibliography{jos}

\begin{thebibliography}{}
\expandafter\ifx\csname url\endcsname\relax
  \def\url#1{\texttt{#1}}\fi
\expandafter\ifx\csname urlprefix\endcsname\relax\def\urlprefix{URL }\fi
\expandafter\ifx\csname href\endcsname\relax
  \def\href#1#2{#2} \def\path#1{#1}\fi

\bibitem{Albacete:2013ei}
  J.~L.~Albacete {\it et al.},
  Int.\ J.\ Mod.\ Phys.\ E {\bf 22} (2013) 1330007
  [arXiv:1301.3395 [hep-ph]].

\bibitem{Albacete:2016veq}
  J.~L.~Albacete {\it et al.},
  Int.\ J.\ Mod.\ Phys.\ E {\bf 25} (2016) no.9,  1630005
  [arXiv:1605.09479 [hep-ph]].


\bibitem{Eskola:2009uj}
  K.~J.~Eskola, H.~Paukkunen and C.~A.~Salgado,
  JHEP {\bf 0904} (2009) 065
  [arXiv:0902.4154 [hep-ph]].

\bibitem{deFlorian:2011fp}
  D.~de Florian, R.~Sassot, P.~Zurita and M.~Stratmann,
  Phys.\ Rev.\ D {\bf 85} (2012) 074028
  [arXiv:1112.6324 [hep-ph]].

\bibitem{Schienbein:2009kk}
  I.~Schienbein, J.~Y.~Yu, K.~Kovarik, C.~Keppel, J.~G.~Morfin, F.~Olness and J.~F.~Owens,
  Phys.\ Rev.\ D {\bf 80} (2009) 094004
  [arXiv:0907.2357 [hep-ph]].

\bibitem{ConesadelValle:2009vp}
  Z.~Conesa del Valle,
  Eur.\ Phys.\ J.\ C {\bf 61} (2009) 729
  [arXiv:0903.1432 [hep-ex]].

\bibitem{Catani:2007vq}
  S.~Catani and M.~Grazzini,
  Phys.\ Rev.\ Lett.\  {\bf 98} (2007) 222002
  [hep-ph/0703012].

\bibitem{Catani:2009sm}
  S.~Catani, L.~Cieri, G.~Ferrera, D.~de Florian and M.~Grazzini,
  Phys.\ Rev.\ Lett.\  {\bf 103} (2009) 082001
  [arXiv:0903.2120 [hep-ph]].


\bibitem{Chatrchyan:2014csa}
  S.~Chatrchyan {\it et al.} [CMS Collaboration],
  JHEP {\bf 1503} (2015) 022
  [arXiv:1410.4825 [nucl-ex]].

\bibitem{Ru:2014yma}
  P.~Ru, B.~W.~Zhang, L.~Cheng, E.~Wang and W.~N.~Zhang,
  J.\ Phys.\ G {\bf 42} (2015) no.8,  085104
  [arXiv:1412.2930 [nucl-th]].

\bibitem{Ru:2015pfa}
  P.~Ru, B.~W.~Zhang, E.~Wang and W.~N.~Zhang,
  Eur.\ Phys.\ J.\ C {\bf 75} (2015) no.9,  426
  [arXiv:1505.08106 [nucl-th]].

\bibitem{Kulagin:2004ie}
  S.~A.~Kulagin and R.~Petti,
  Nucl.\ Phys.\ A {\bf 765} (2006) 126
  [hep-ph/0412425].

\bibitem{Kulagin:2014vsa}
  S.~A.~Kulagin and R.~Petti,
  Phys.\ Rev.\ C {\bf 90} (2014) no.4,  045204
  [arXiv:1405.2529 [hep-ph]].

\bibitem{Ru:2016wfx}
  P.~Ru, S.~A.~Kulagin, R.~Petti and B.~W.~Zhang,
  arXiv:1608.06835 [nucl-th].

\bibitem{Khachatryan:2015hha}
  V.~Khachatryan {\it et al.} [CMS Collaboration],
  Phys.\ Lett.\ B {\bf 750} (2015) 565
  [arXiv:1503.05825 [nucl-ex]].


\bibitem{Alekhin:2015cza}
  S.~Alekhin, J.~Bluemlein, S.~Moch and R.~Placakyte,
  arXiv:1508.07923 [hep-ph].


\bibitem{Gao:2013xoa}
  J.~Gao {\it et al.},
  Phys.\ Rev.\ D {\bf 89} (2014) no.3,  033009
  [arXiv:1302.6246 [hep-ph]].



\end{thebibliography}







\end{document}